\documentclass{aastex}          
\usepackage{spr-astr-addons}    

\begin{document}
\overfullrule=0pt
\title{Possible Relations between Brightest Central Galaxies and Their Host Galaxies Clusters and Groups}

\shorttitle{Possible Relations between Brightest Central Galaxies and Their Host Galaxies Clusters and Groups}
\shortauthors{Samir et al. }

\author{R. M. Samir\altaffilmark{1}} 
\and 
\author{A. A. Shaker\altaffilmark{1}}
\affil{rasha.samir$@$nriag.sci.eg, shaker@nriag.sci.eg}

\altaffiltext{1}{National Research Institute of Astronomy and Geophysics (NRIAG), 11421 Helwan, Cairo, Egypt}

\maketitle 
	
\begin{abstract}
The \textit{r}-band of the Sloan Digital Sky Survey (SDSS) for 17,924 Brightest Cluster Galaxies (BCGs) in clusters and groups within $0.02 \leqslant z \leqslant 0.20$ have been used to study possible environmental relations affecting the nature of these galaxies. We find a correlation between BCGs physical properties (the effective radius, absolute magnitude and central velocity dispersion (\textit{$\sigma_{0}$})) and host groups and clusters velocity dispersion (\textit{$\sigma _{cl}$}). This type of relations suggests that the most massive groups or clusters host larger central galaxies. On the other hand, the \textit{$\sigma_{0}$}/\textit{$\sigma _{cl}$} ratio as a function of \textit{$\sigma_{cl}$} is consistent with \cite{Sohn2020}.

\end{abstract}


\section{Introduction}
BCGs are considered as one from the interesting objects that can explain some unsolved problems in studying galaxies. They are large, bright early type galaxies lying at centers of most galaxies groups and clusters.  

There are several unanswered questions related to BCGs nature. The principal mechanism for BCGs evolution is feedback rather than merging \citep[e.g.][]{Ascaso2011}. BCGs properties may also be affected by their cluster halos mass at low redshift. BCGs Fundamental Plane (FP) in general follow the same FP of normal clusters \cite{Samir2011}, but BCGs FP and scaling relations are not the same as those of isolated galaxies and the FP is a waveband dependent \cite{Samir2020}. 

Previous studies found correlations between BCGs and their clusters properties such as cluster mass, richness and X-ray luminosity \citep[e.g.][]{Stott2011, Takey2014, Cerulo2019}. 
Understanding BCGs nature needs to study well environmental effects which affect on their physical properties \citep[e.g.][]{Shaker98, Samir2011, Samir2016}. 

In this work we present the sample and focus on examining the properties of the BCGs and \textit{$\sigma _{cl}$} relation. We investigate BCGs physical properties as the effective radii, absolute magnitude, \textit{$\sigma _o$} and  \textit{$\sigma _{cl}$}.

This paper is organized as follows; Section 2 describes the selection of our BCGs sample, while in Section 3 we introduce and discuss our results and in Section 4 our main conclusions are summarized. In this paper we will adopt the cosmological parameters $\Omega_m$ = 0.3, $\Omega_\Lambda$ = 0.7, and 
Hubble constant H$_o$ =70 km s$^{-1}$ Mpc$^{-1}$.


\section{Galaxy Sample}

BCGs and Brightest Groups Galaxies (BGGs) have been selected using \cite{Wang2014} Catalog. A sample of 17,924 Galaxies in the r-band of the Sloan Digital Sky Survey (SDSS) in clusters and groups within $0.02 \leqslant z\leqslant 0.20$ have been constructed.  All parameters in our sample are obtained and updated from the SDSS-DR16. Physical properties of BCGs and BGGs (the effective radius, absolute magnitude and central velocity dispersion (\textit{$\sigma _o$})) are derived and corrected as in \cite{Samir2020}. The velocity dispersion of host groups and clusters (\textit{$\sigma _{cl}$}) are taken from \cite{Wang2014}.

\section {Results and Discussions}
\subsection{Correlations between BCGs and environments}
In this section, BCGs - environments relations have been studied. Figure 1 shows that Most of clusters and groups have \textit{$\sigma _{cl}$} higher than 300 km s$^{-1}$.

\begin{figure}[h]
\includegraphics[scale=0.30]{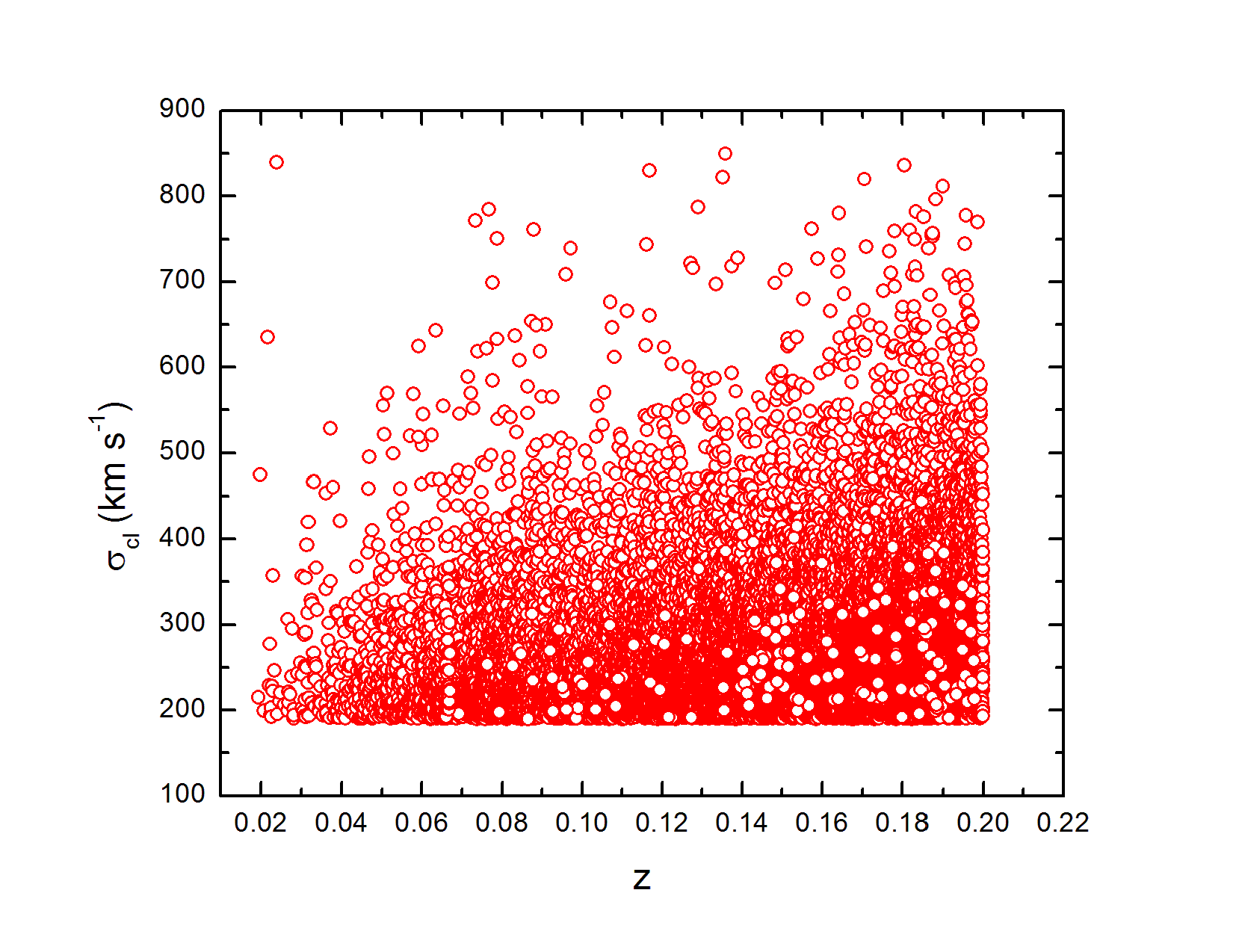}                           
		\caption{\textit{$\sigma _{cl}$} - redshift relation.}
\end{figure}

Figure 2 describes BCGs \textit{r}-band effective radii distribution with their spectroscopic redshift.

\begin{figure}[h]
\includegraphics[scale=0.30] {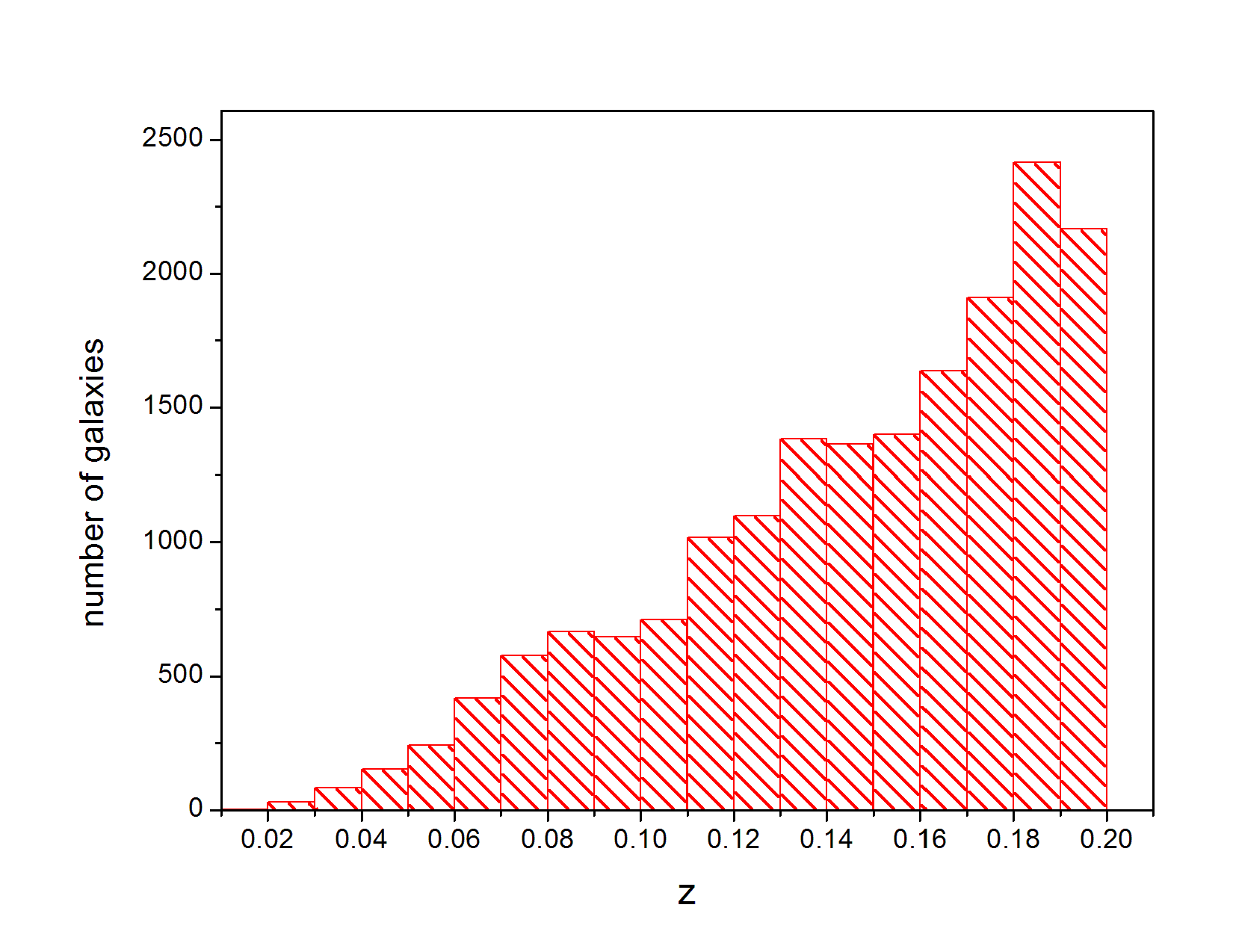}                           
		\caption{ BCGs distribution with their spectroscopic redshifts.}
\end{figure}

Figure 3(a) shows the \textit{r}-band effective radii (\textit{R$_e$}) of BCGs with respects to clusters and groups redshift. It is clear that galaxies are having larger \textit{R$_e$} as the redshift of their host environment increases. Figure 3(b) shows \textit{R$_e$} of the BCGs as a function of \textit{$\sigma _{cl}$}. Error bars represent the average values of \textit{$\sigma _{cl}$} in every \textit{R$_e$} bins with errors. A tight correlation between BCGs \textit{R$_e$} and \textit{$\sigma _{cl}$} of their clusters or groups, is existed. The most massive clusters and groups host central galaxies with larger values of \textit{R$_e$}.

\begin{figure}[h]
\includegraphics[scale=0.30] {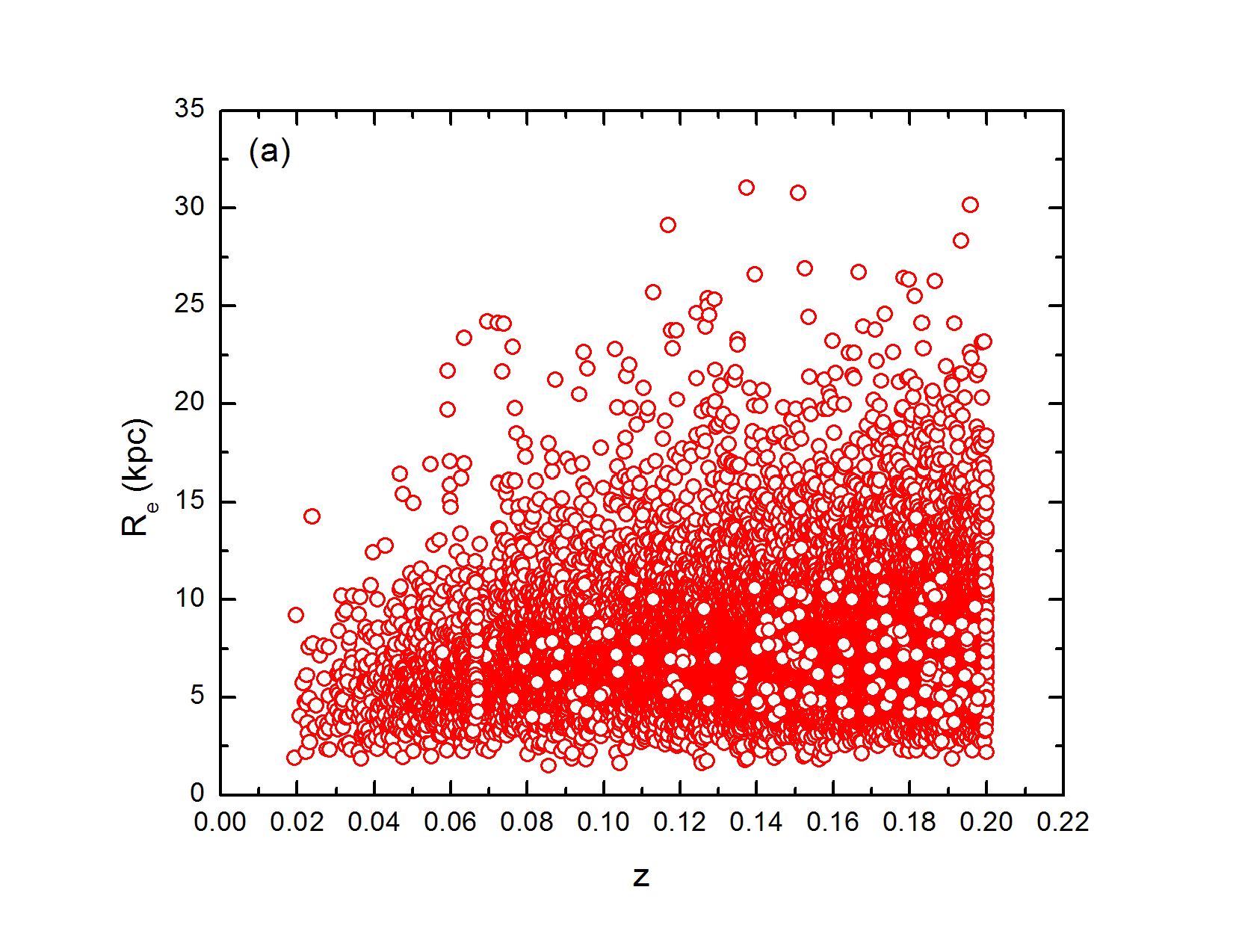} 
\includegraphics[scale=0.30] {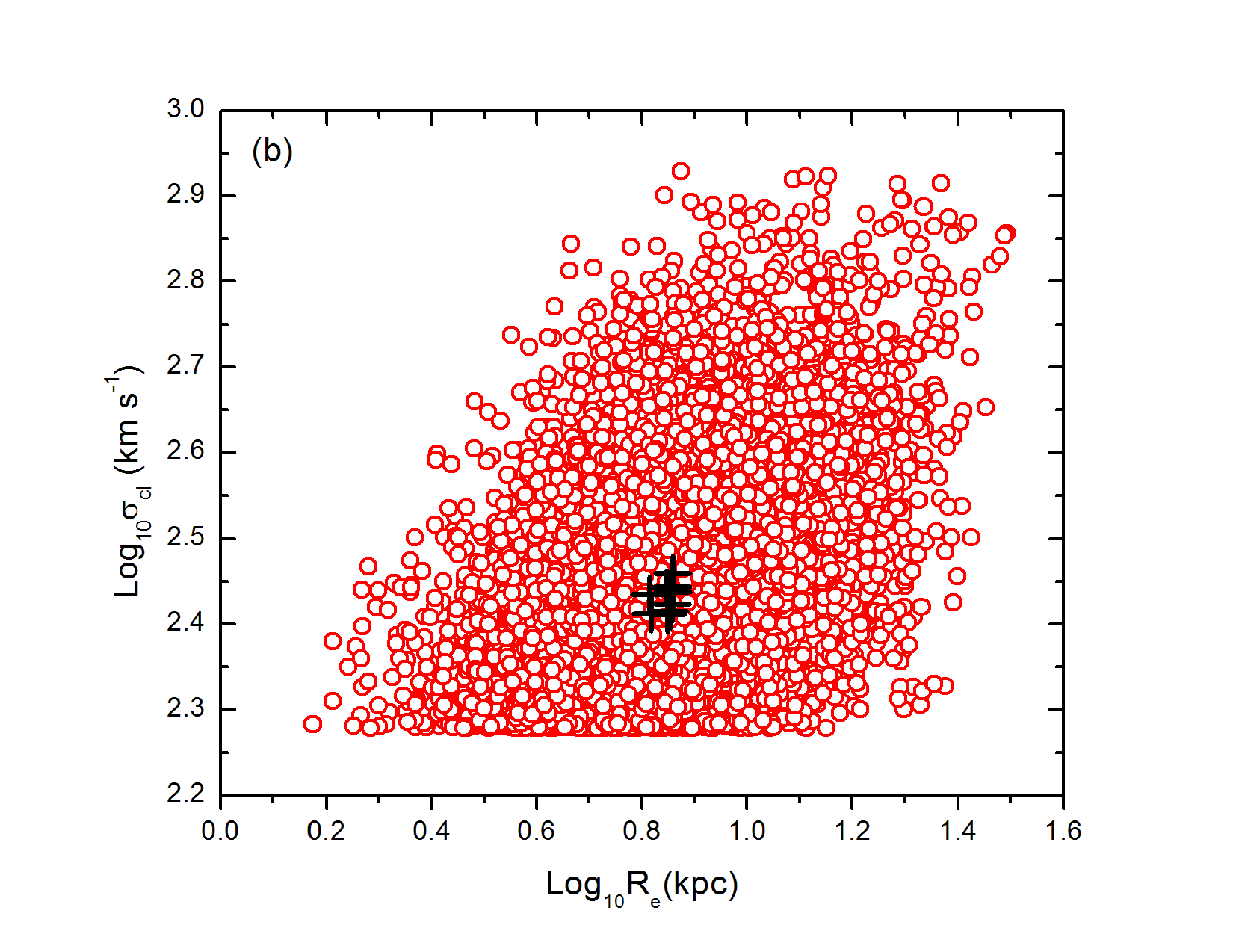}                           
		\caption{BCGs \textit{r}-band effective radii with (a) redshift and (b) \textit{$\sigma _{cl}$} of host clusters and groups. Error bars represent the average values of \textit{$\sigma _{cl}$} in every \textit{R$_e$} bin with errors.}
\end{figure}

Figure 4(a) shows the \textit{r}-band BCGs absolute magnitude as a function of their host redshift. While BCGs are very bright (M$_r$ $<$ -22) across redshift range, low redshift clusters and groups (z $\leq$ 0.04) hosts less luminous BCGs. These low redshift systems have also lower values of \textit{$\sigma _{cl}$} as clear in Figure 1. Figure 4(b) shows the \textit{r}-band absolute magnitude of BCGs as a function of \textit{$\sigma _{cl}$}. It is clear from this figure that clusters or groups with higher velocity dispersions host the brighter central galaxies which is consistent with previous studies \citep[see][]{Lin2004, Wen2012, Wen2018}.

\begin{figure}[h]
\includegraphics[scale=0.30] {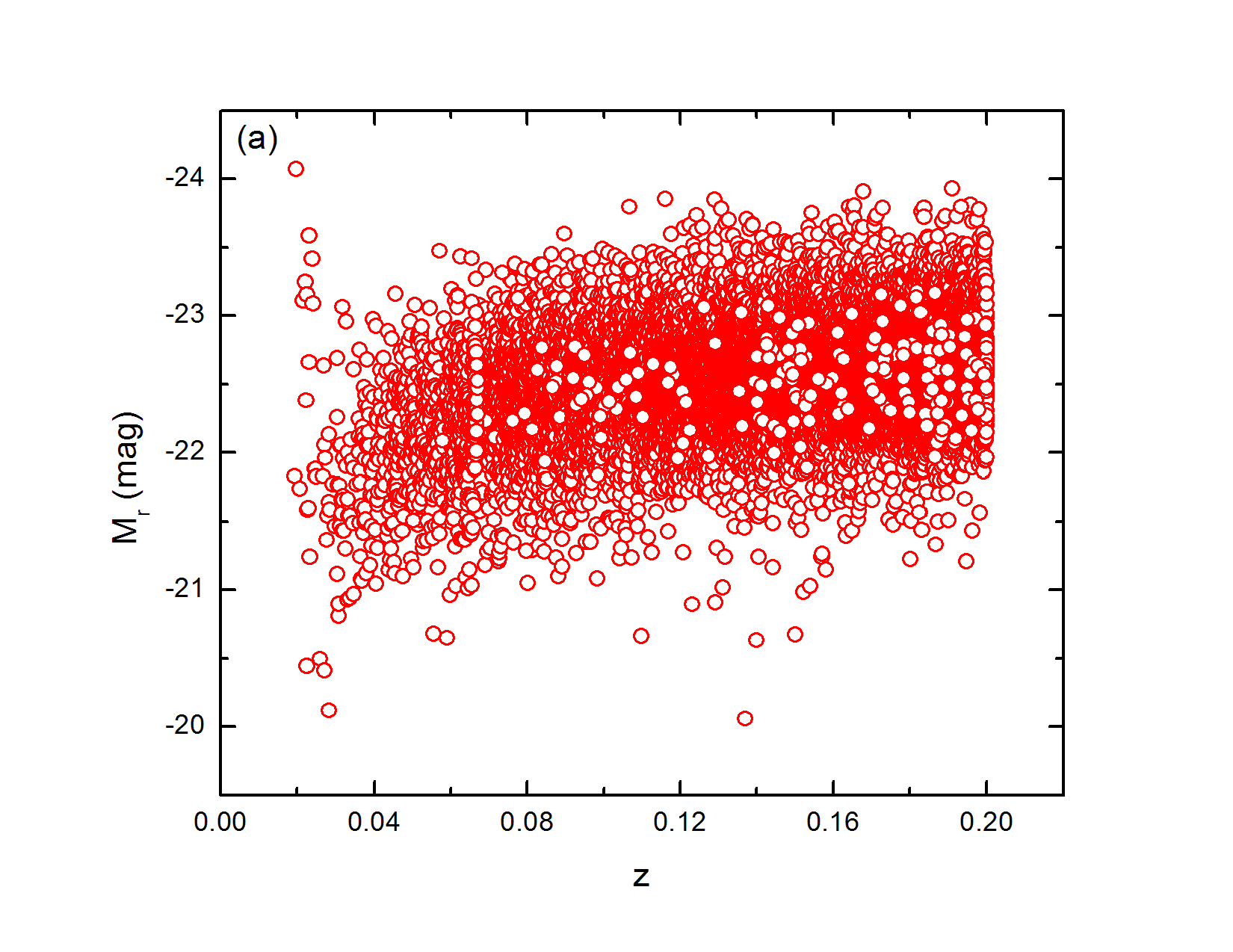} 
\includegraphics[scale=0.30] {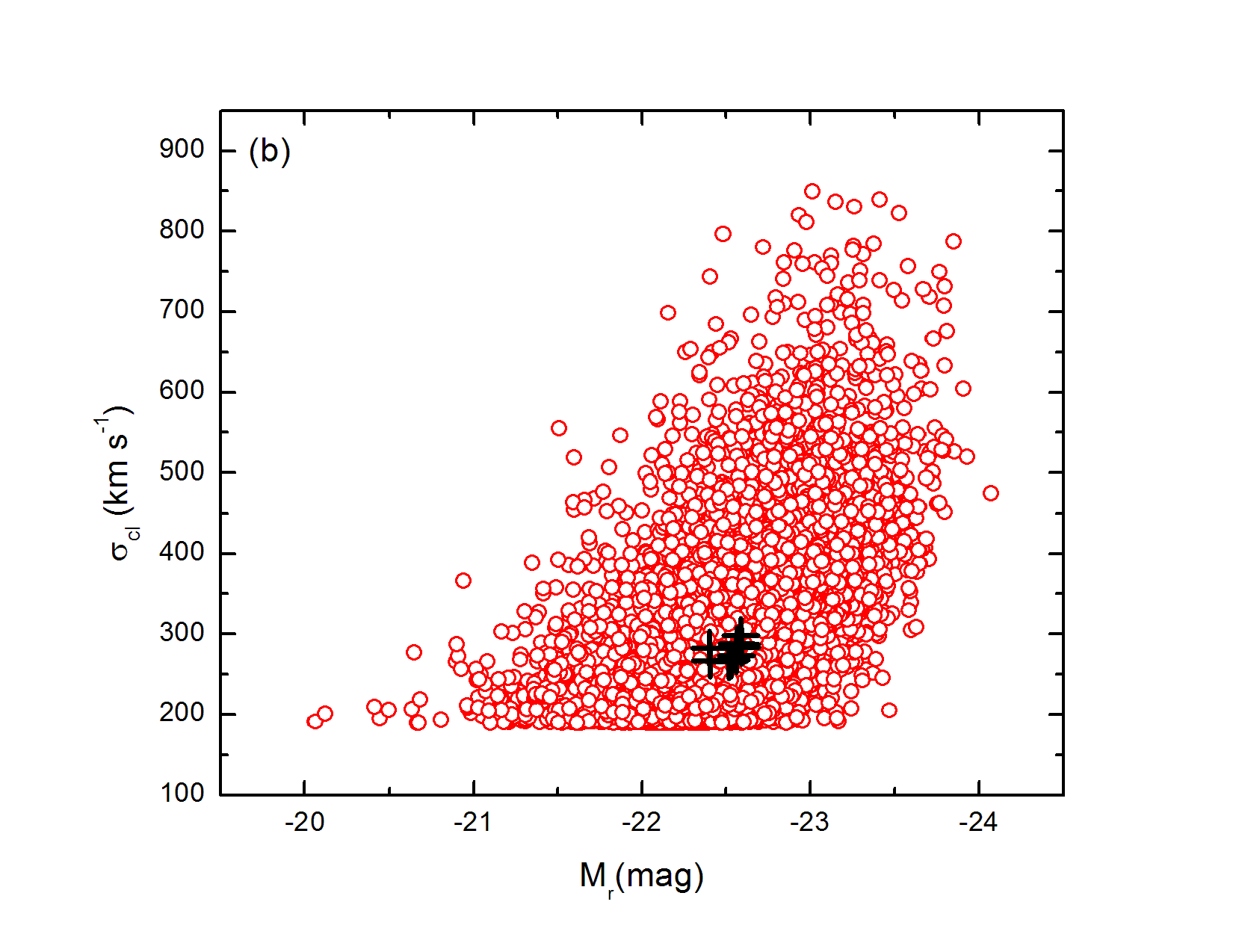}                           
		\caption{BCGs \textit{r}-band absolute magnitude with respect to (a) redshift and (b) \textit{$\sigma _{cl}$} of host clusters and groups. Error bars represent the average values of \textit{$\sigma _{cl}$} in every M$_r$ bin with errors.}
\end{figure}

Figure 5(a) shows BCGs \textit{$\sigma _o$} in \textit{r}-band with their host redshift. In general, values of \textit{$\sigma _o$} increase along the redsift range. In Figure 5(b), we plot BCGs   \textit{$\sigma _o$} as a function of \textit{$\sigma _{cl}$} of their host. This figure shows that there is a correlation between BCGs  \textit{$\sigma _o$} and \textit{$\sigma _{cl}$} of their host environment. We can say that BCGs \textit{$\sigma _o$} depends on its host cluster or group mass. The observed relation between \textit{$\sigma _o$} and \textit{$\sigma _{cl}$} provides an important test for the formation models of galaxies and clusters.

\begin{figure}[h]
\includegraphics[scale=0.30] {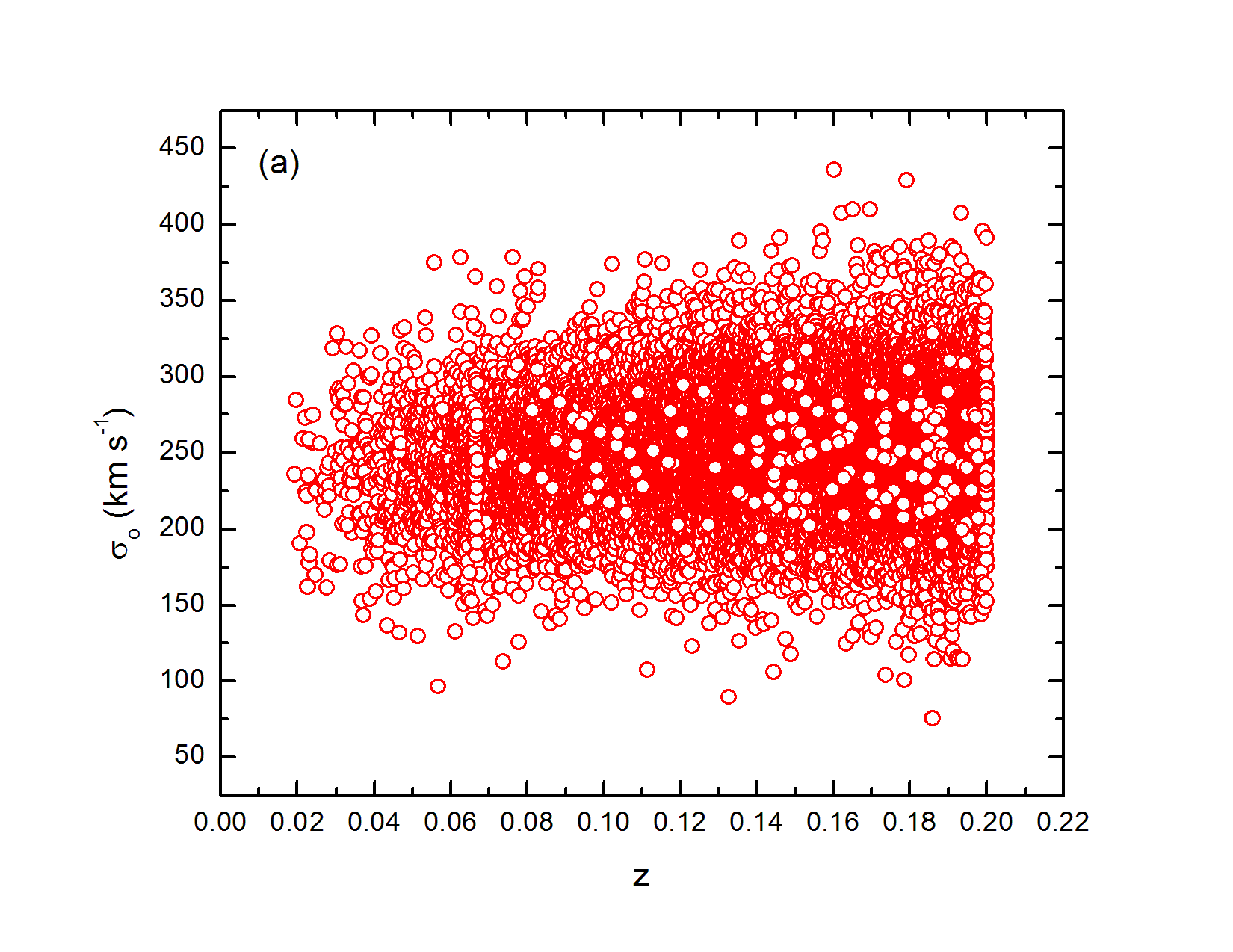} 
\includegraphics[scale=0.30] {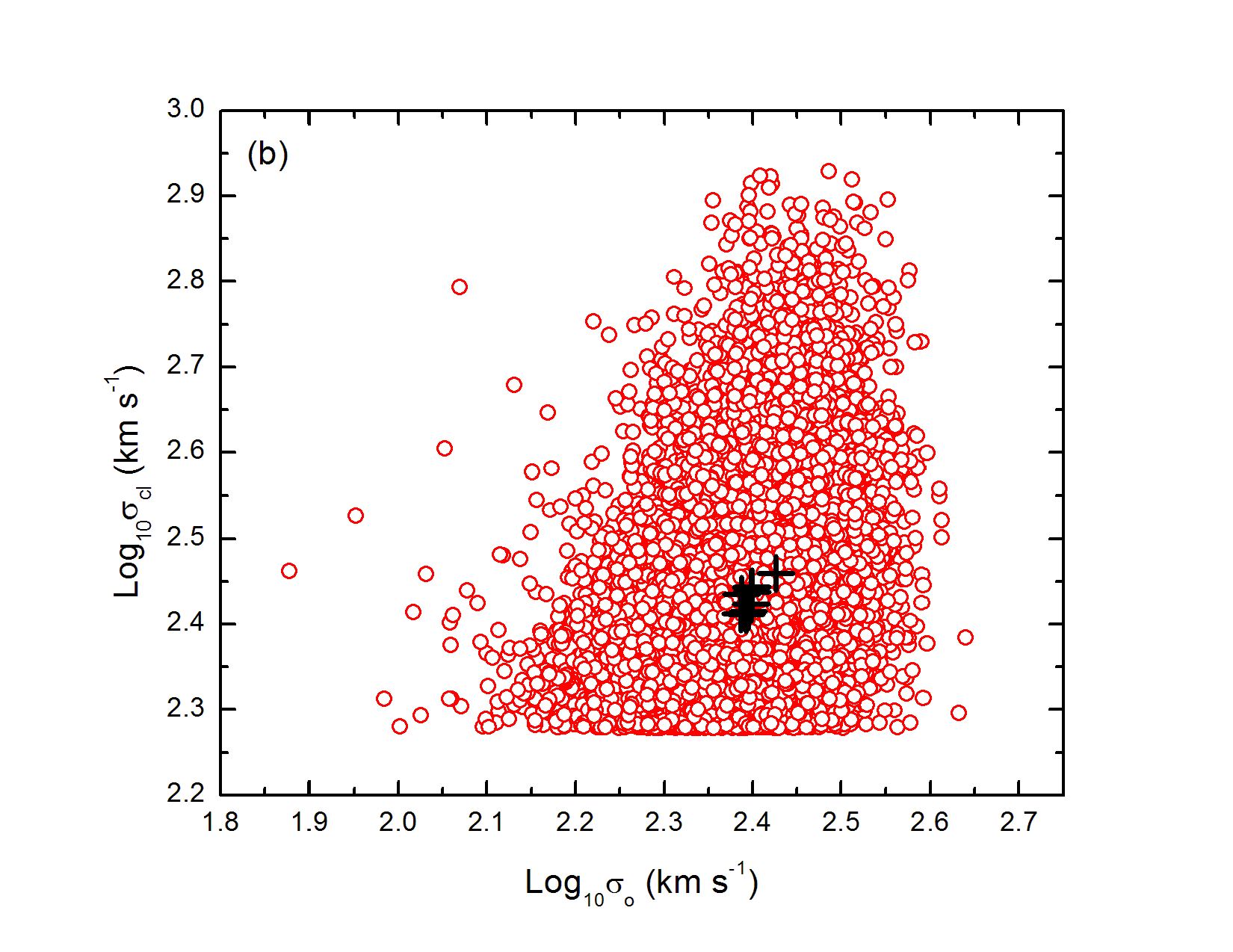}                           
		\caption{BCGs \textit{$\sigma _o$} in \textit{r}-band with respect to (a) redshift and (b) \textit{$\sigma _{cl}$} 	of host clusters and groups. Error bars represent the average values of log$_{10}$ \textit{$\sigma _{cl}$} for log$_{10}$ \textit{$\sigma _o$} bins with errors.}
\end{figure}

\subsection{Comparison with simulations}

\begin{figure}[h]
\includegraphics[scale=0.30]{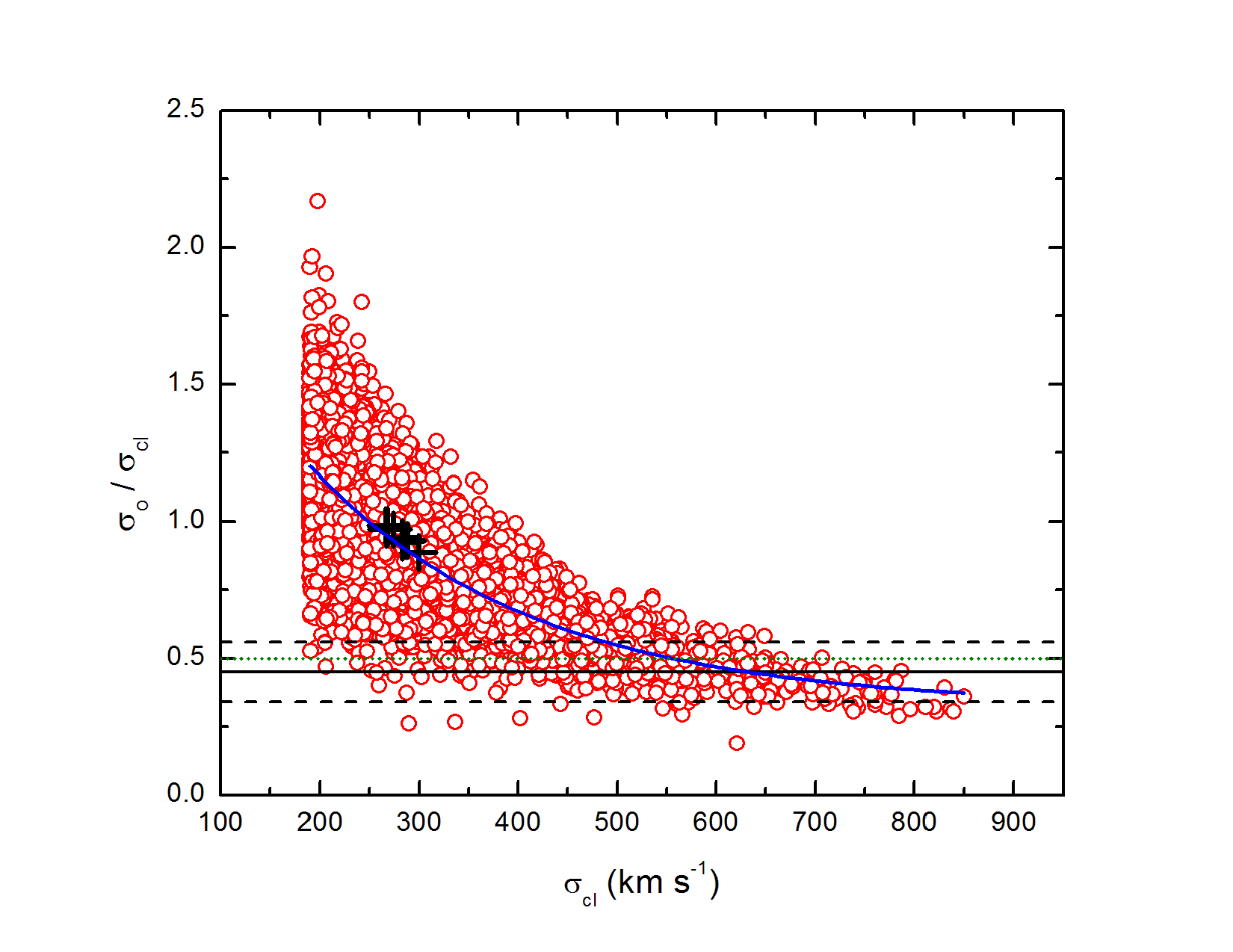}                           
		\caption{BCGs $\sigma_o$ in \textit{r}-band with respect to \textit{$\sigma_{cl}$} of host clusters and groups as a function of \textit{$\sigma_{cl}$}. Error bars represent the 
		average values of \textit{$\sigma_o$}/\textit{$\sigma_{cl}$} in different \textit{$\sigma_{cl}$} bins with errors. The blue solid line shows the best exponential fit to our sample. 
		The black solid and dashed lines show the predicted relation and the 1\textit{$\sigma$} deviation from simulation done by \cite{Dolag2010},respectively. The green dotted line is the predicted relation from simulations done by \cite{Remus2017}.}
\end{figure}

We found that BCGs \textit{$\sigma_o$} correlates with \textit{$\sigma_{cl}$} which represents a cluster and group halo mass proxy. Here we compare the observed relation with predictions from numerical simulations by \cite{Dolag2010} and \cite{Remus2017}. \cite{Dolag2010} performed their simulation taking into account heating by a UV background, radiative cooling, star
formation and feedback. 44 clusters have been identified with more than 20 satellite galaxies. Some star particles are identified which are not bound to any subhalo within the cluster potential. Those star particles are either a diffuse stellar component (DSC) or the BCGs stellar component (cD galaxies). They separated those two components and calculated velocity dispersions for them .

Figure 6 shows that the relation between BCGs \textit{$\sigma_o$} and  \textit{$\sigma_{cl}$} decreases a function of  \textit{$\sigma_{cl}$} which is also consistent with previous result from \cite{Sohn2020}. They studied the relation between 227 BCGs and their host clusters in the redshift range 0.02 $<$ z $<$ 0.30. They found a tight relation between BCG velocity dispersion and cluster velocity dispersion. They found that the ratio of BCGs \textit{$\sigma_o$} and \textit{$\sigma_{cl}$} decreases as a function of cluster velocity dispersion. On the other side, \cite{Dolag2010} simulation suggests a constant relation between \textit{$\sigma_o$} and \textit{$\sigma_{cl}$} along a wide range of \textit{$\sigma_{cl}$}. Also numerical simulations of \cite{Remus2017} represents the same constant relation. Error bars represent the average values of \textit{$\sigma_o$}/\textit{$\sigma_{cl}$} in different \textit{$\sigma_{cl}$} bins with errors.

In the simulation of \cite{Dolag2010}, the velocity dispersion of BCGs, DSC, and cluster galaxies are well correlated with virial mass of cluster halo. We can say that the observed \textit{$\sigma_o$}/\textit{$\sigma_{cl}$} ratio depends on cluster and group mass which suggests that the mass fraction associated with brightest central galaxies indicate the evolution of brightest central galaxies and their host halos.

\section{Summary and Conclusions}

The \textit{r}-band of the Sloan Digital Sky Survey for 17,924 Brightest Cluster Galaxies in clusters and groups within $0.02 \leqslant z \leqslant 0.20$  have been used to study possible environmental relations in the formation of these galaxies. All parameters in our sample are obtained and updated from the SDSS-DR16. The relation between BCGs effective radii  with velocity dispersions of their host clusters and groups is tight. As a result, the most massive clusters and groups host larger central galaxies. Also the \textit{r}-band absolute magnitude correlate with \textit{$\sigma _{cl}$}. This indicates that clusters and groups with higher velocity dispersions host the more bright central galaxies, that is in agreement with other studies \citep[see][]{Lin2004, Wen2012, Wen2018}. We find that BCGs \textit{$\sigma _o$} depend on its host group or cluster mass and \textit{$\sigma _o$} correlates tightly with \textit{$\sigma_{cl}$}. This relation suggests that BCGs \textit{$\sigma _o$} can be used as a tracer for group or cluster mass, which is in agreement with \cite{Shaker98}. On the other hand, the \textit{$\sigma_o$}/\textit{$\sigma_{cl}$} ratio decreases as \textit{$\sigma_{cl}$} increases which is consistent with \cite{Sohn2020}. Also \textit{$\sigma_o$}/\textit{$\sigma_{cl}$} ratio is different than \cite{Dolag2010} theoretical predictions, which suggests two different scenarios for BCGs growth in lower and massive host groups and clusters. More large scale simulations studies of \textit{$\sigma_o$} and \textit{$\sigma_{cl}$} is useful for studying BCGs environmental dependence.

\acknowledgments
Funding for the Sloan Digital Sky Survey (SDSS) has been provided by the Alfred P. Sloan Foundation, the Participating Institutions, the National Aeronautics and Space Administration, the National Science Foundation, the U.S. Department of Energy, the Japanese Monbukagakusho, and the Max Planck Society. The SDSS Web site is http://www.sdss.org/.

The SDSS is managed by the Astrophysical Research Consortium (ARC) for the Participating Institutions. The Participating Institutions are The University of Chicago, Fermilab, the Institute for Advanced Study, the Japan Participation Group, The Johns Hopkins University, Los Alamos National Laboratory, the Max-Planck-Institute for Astronomy (MPIA), the Max-Planck-Institute for Astrophysics (MPA), New Mexico State University, University of Pittsburgh, Princeton University, the United States Naval Observatory, and the University of Washington. 




\bibliographystyle{spr-mp-nameyear-cnd}
\bibliography{ref.bib}

\begin{thebibliography}{14}
\ifx \bisbn   \undefined \def \bisbn  #1{ISBN #1}\fi
\ifx \binits  \undefined \def \binits#1{#1} \fi
\ifx \bauthor  \undefined \def \bauthor#1{#1} \fi
\ifx \batitle  \undefined \def \batitle#1{#1} \fi
\ifx \bjtitle  \undefined \def \bjtitle#1{#1}\fi
\ifx \bvolume  \undefined \def \bvolume#1{\textbf{#1}}\fi
\ifx \byear  \undefined \def \byear#1{#1} \fi
\ifx \bissue  \undefined \def \bissue#1{#1} \fi
\ifx \bfpage  \undefined \def \bfpage#1{#1} \fi
\ifx \blpage  \undefined \def \blpage #1{#1} \fi
\ifx \burl  \undefined \def \burl#1{\textsf{#1}} \fi
\ifx \doiurl  \undefined \def \doiurl#1{\textsf{#1}} \fi
\ifx \betal  \undefined \def \betal{\textit{et al.}} \fi
\ifx \binstitute  \undefined \def \binstitute#1{#1} \fi
\ifx \binstitutionaled  \undefined \def \binstitutionaled#1{#1} \fi
\ifx \bctitle  \undefined \def \bctitle#1{#1} \fi
\ifx \beditor  \undefined \def \beditor#1{#1} \fi
\ifx \bpublisher  \undefined \def \bpublisher#1{#1} \fi
\ifx \bbtitle  \undefined \def \bbtitle#1{#1} \fi
\ifx \bedition  \undefined \def \bedition#1{#1} \fi
\ifx \bseriesno  \undefined \def \bseriesno#1{#1} \fi
\ifx \blocation  \undefined \def \blocation#1{#1} \fi
\ifx \bsertitle  \undefined \def \bsertitle#1{#1} \fi
\ifx \bsnm \undefined \def \bsnm#1{#1} \fi
\ifx \bsuffix \undefined \def \bsuffix#1{#1} \fi
\ifx \bparticle \undefined \def \bparticle#1{#1} \fi
\ifx \barticle \undefined \def \barticle#1{#1} \fi
\ifx \bconfdate \undefined \def \bconfdate #1{#1} \fi
\ifx \botherref \undefined \def \botherref #1{#1} \fi
\ifx \url \undefined \def \url#1{\textsf{#1}} \fi
\ifx \bchapter \undefined \def \bchapter#1{#1} \fi
\ifx \bbook \undefined \def \bbook#1{#1} \fi
\ifx \bcomment \undefined \def \bcomment#1{#1} \fi
\ifx \oauthor \undefined \def \oauthor#1{#1} \fi
\ifx \citeauthoryear \undefined \def \citeauthoryear#1{#1} \fi
\ifx \endbibitem  \undefined \def \endbibitem {}\fi
\ifx \bconflocation  \undefined \def \bconflocation#1{#1} \fi
\ifx \arxivurl  \undefined \def \arxivurl#1{\textsf{#1}} \fi

\bibitem[\protect\citeauthoryear{{Ascaso} et~al.}{2011}]{Ascaso2011}
\begin{barticle}
\bauthor{\bsnm{{Ascaso}}, \binits{B.}},
\bauthor{\bsnm{{Aguerri}}, \binits{J.A.L.}},
\bauthor{\bsnm{{Varela}}, \binits{J.}},
\bauthor{\bsnm{{Cava}}, \binits{A.}},
\bauthor{\bsnm{{Bettoni}}, \binits{D.}},
\bauthor{\bsnm{{Moles}}, \binits{M.}},
\bauthor{\bsnm{{D'Onofrio}}, \binits{M.}}:
\bjtitle{\apj}
\bvolume{726}(\bissue{2}),
\bfpage{69}
(\byear{2011}).
\arxivurl{1007.3264}.
doi:\doiurl{10.1088/0004-637X/726/2/69}
\end{barticle}
\endbibitem

\bibitem[\protect\citeauthoryear{{Cerulo} et~al.}{2019}]{Cerulo2019}
\begin{botherref}
\oauthor{\bsnm{{Cerulo}}, \binits{P.}},
\oauthor{\bsnm{{Orellana}}, \binits{G.A.}},
\oauthor{\bsnm{{Covone}}, \binits{G.}}:
\mnras
(2019).
\arxivurl{1905.12117}.
doi:\doiurl{10.1093/mnras/stz1495}
\end{botherref}
\endbibitem

\bibitem[\protect\citeauthoryear{{Dolag} et~al.}{2010}]{Dolag2010}
\begin{barticle}
\bauthor{\bsnm{{Dolag}}, \binits{K.}},
\bauthor{\bsnm{{Murante}}, \binits{G.}},
\bauthor{\bsnm{{Borgani}}, \binits{S.}}:
\bjtitle{\mnras}
\bvolume{405}(\bissue{3}),
\bfpage{1544}
(\byear{2010}).
\arxivurl{0911.1129}.
doi:\doiurl{10.1111/j.1365-2966.2010.16583.x}
\end{barticle}
\endbibitem

\bibitem[\protect\citeauthoryear{{Remus} et~al.}{2017}]{Remus2017}
\begin{barticle}
\bauthor{\bsnm{{Remus}}, \binits{R.-S.}},
\bauthor{\bsnm{{Dolag}}, \binits{K.}},
\bauthor{\bsnm{{Hoffmann}}, \binits{T.}}:
\bjtitle{Galaxies}
\bvolume{5}(\bissue{3}),
\bfpage{49}
(\byear{2017}).
\arxivurl{1709.02393}.
doi:\doiurl{10.3390/galaxies5030049}
\end{barticle}
\endbibitem

\bibitem[\protect\citeauthoryear{{Samir} et~al.}{2020}]{Samir2020}
\begin{barticle}
\bauthor{\bsnm{{Samir}}, \binits{R.M.}},
\bauthor{\bsnm{{Takey}}, \binits{A.}},
\bauthor{\bsnm{{Shaker}}, \binits{A.A.}}:
\bjtitle{\apss}
\bvolume{365}(\bissue{8}),
\bfpage{142}
(\byear{2020}).
doi:\doiurl{10.1007/s10509-020-03857-8}
\end{barticle}
\endbibitem

\bibitem[\protect\citeauthoryear{{Samir} et~al.}{2011}]{Samir2011}
\begin{barticle}
\bauthor{\bsnm{{Samir}}, \binits{R.M.}},
\bauthor{\bsnm{{Reda}}, \binits{F.M.}},
\bauthor{\bsnm{{Shaker}}, \binits{A.A.}},
\bauthor{\bsnm{{Osman}}, \binits{A.M.I.}},
\bauthor{\bsnm{{Amin}}, \binits{M.Y.}}:
\bjtitle{NRIAG Journal of Astronomy and Geophysics}
\bvolume{5},
\bfpage{237}
(\byear{2011}).
doi:\doiurl{10.1016/j.nrjag.2016.06.004}
\end{barticle}
\endbibitem

\bibitem[\protect\citeauthoryear{{Samir} et~al.}{2016}]{Samir2016}
\begin{barticle}
\bauthor{\bsnm{{Samir}}, \binits{R.M.}},
\bauthor{\bsnm{{Reda}}, \binits{F.M.}},
\bauthor{\bsnm{{Shaker}}, \binits{A.A.}},
\bauthor{\bsnm{{Osman}}, \binits{A.M.I.}},
\bauthor{\bsnm{{Amin}}, \binits{M.Y.}}:
\bjtitle{NRIAG Journal of Astronomy and Geophysics}
\bvolume{5},
\bfpage{277}
(\byear{2016}).
doi:\doiurl{10.1016/j.nrjag.2016.06.004}
\end{barticle}
\endbibitem

\bibitem[\protect\citeauthoryear{{Shaker} et~al.}{1998}]{Shaker98}
\begin{barticle}
\bauthor{\bsnm{{Shaker}}, \binits{A.A.}},
\bauthor{\bsnm{{Longo}}, \binits{G.}},
\bauthor{\bsnm{{Merluzzi}}, \binits{P.}}:
\bjtitle{Astronomische Nachrichten}
\bvolume{319},
\bfpage{167}
(\byear{1998}).
\arxivurl{astro-ph/9804087}.
doi:\doiurl{10.1002/asna.2123190305}
\end{barticle}
\endbibitem

\bibitem[\protect\citeauthoryear{{Sohn} et~al.}{2020}]{Sohn2020}
\begin{barticle}
\bauthor{\bsnm{{Sohn}}, \binits{J.}},
\bauthor{\bsnm{{Geller}}, \binits{M.J.}},
\bauthor{\bsnm{{Diaferio}}, \binits{A.}},
\bauthor{\bsnm{{Rines}}, \binits{K.J.}}:
\bjtitle{\apj}
\bvolume{891}(\bissue{2}),
\bfpage{129}
(\byear{2020}).
\arxivurl{1910.11192}.
doi:\doiurl{10.3847/1538-4357/ab6e6a}
\end{barticle}
\endbibitem

\bibitem[\protect\citeauthoryear{{Stott} et~al.}{2011}]{Stott2011}
\begin{barticle}
\bauthor{\bsnm{{Stott}}, \binits{J.P.}},
\bauthor{\bsnm{{Collins}}, \binits{C.A.}},
\bauthor{\bsnm{{Burke}}, \binits{C.}},
\bauthor{\bsnm{{Hamilton-Morris}}, \binits{V.}},
\bauthor{\bsnm{{Smith}}, \binits{G.P.}}:
\bjtitle{\mnras}
\bvolume{414}(\bissue{1}),
\bfpage{445}
(\byear{2011}).
\arxivurl{1101.4652}.
doi:\doiurl{10.1111/j.1365-2966.2011.18404.x}
\end{barticle}
\endbibitem

\bibitem[\protect\citeauthoryear{{Takey}}{2014}]{Takey2014}
\begin{botherref}
\oauthor{\bsnm{{Takey}}, \binits{A.}}:
"The XMM-Newton/SDSS Galaxy Cluster Survey" PhD thesis, Potsdam University
  (date: 16.12.2013, arXiv: 1407.6566)
(2014).
\arxivurl{1407.6566}
\end{botherref}
\endbibitem

\bibitem[\protect\citeauthoryear{{Wang} et~al.}{2014}]{Wang2014}
\begin{barticle}
\bauthor{\bsnm{{Wang}}, \binits{L.}},
\bauthor{\bsnm{{Yang}}, \binits{X.}},
\bauthor{\bsnm{{Shen}}, \binits{S.}},
\bauthor{\bsnm{{Mo}}, \binits{H.J.}},
\bauthor{\bsnm{{van den Bosch}}, \binits{F.C.}},
\bauthor{\bsnm{{Luo}}, \binits{W.}},
\bauthor{\bsnm{{Wang}}, \binits{Y.}},
\bauthor{\bsnm{{Lau}}, \binits{E.T.}},
\bauthor{\bsnm{{Wang}}, \binits{Q.D.}},
\bauthor{\bsnm{{Kang}}, \binits{X.}},
\bauthor{\bsnm{{Li}}, \binits{R.}}:
\bjtitle{\mnras}
\bvolume{439}(\bissue{1}),
\bfpage{611}
(\byear{2014}).
\arxivurl{1312.7417}.
doi:\doiurl{10.1093/mnras/stt2481}
\end{barticle}
\endbibitem

\bibitem[\protect\citeauthoryear{{Wen} and {Han}}{2018}]{Wen2018}
\begin{barticle}
\bauthor{\bsnm{{Wen}}, \binits{Z.L.}},
\bauthor{\bsnm{{Han}}, \binits{J.L.}}:
\bjtitle{\mnras}
\bvolume{481}(\bissue{3}),
\bfpage{4158}
(\byear{2018}).
\arxivurl{1809.05223}.
doi:\doiurl{10.1093/mnras/sty2533}
\end{barticle}
\endbibitem

\bibitem[\protect\citeauthoryear{{Wen} et~al.}{2012}]{Wen2012}
\begin{barticle}
\bauthor{\bsnm{{Wen}}, \binits{Z.L.}},
\bauthor{\bsnm{{Han}}, \binits{J.L.}},
\bauthor{\bsnm{{Liu}}, \binits{F.S.}}:
\bjtitle{\apjs}
\bvolume{199}(\bissue{2}),
\bfpage{34}
(\byear{2012}).
\arxivurl{1202.6424}.
doi:\doiurl{10.1088/0067-0049/199/2/34}
\end{barticle}
\endbibitem

\end{thebibliography}

\end{document}